\documentclass[10pt,aps,prb,superscriptaddress,twocolumn,floatfix]{revtex4-2}

\usepackage{graphicx}
\usepackage{xcolor}
\usepackage{url} 
\usepackage{hyperref} 
\usepackage{dcolumn}
\usepackage{bm}
\usepackage{multirow}
\usepackage[symbol]{footmisc}
\usepackage{amsmath,amssymb,amsfonts}
\usepackage{subfigure}
\usepackage{soul}

\newcommand{\be}{\begin{eqnarray}}
\newcommand{\ee}{\end{eqnarray}}

\begin{document}

\title{Vector-field control and emergent basal-plane anisotropy of magnetic textures in noncentrosymmetric (Fe$_{0.63}$Ni$_{0.3}$Pd$_{0.07}$)$_3$P}

\author{Victor Ukleev}
\email{victor.ukleev@helmholtz-berlin.de}
\affiliation{Helmholtz-Zentrum Berlin f\"ur Materialien und Energie, D-12489 Berlin, Germany}

\author{Oleg I. Utesov}
\email{utiosov@gmail.com}
\affiliation{Department of Physics, Korea Advanced Institute of Science and Technology, Daejeon 34141, Republic of Korea}
\affiliation{Center for Theoretical Physics of Complex Systems, Institute for Basic Science (IBS), Daejeon 34126, Republic of Korea}

\author{Lorenzo Ubilla}
\author{Chen Luo}
\author{Radu-Marius Abrudan}
\affiliation{Helmholtz-Zentrum Berlin f\"ur Materialien und Energie, D-12489 Berlin, Germany}

\author{Peter Wild}
\affiliation{Heinz Maier-Leibnitz Zentrum (MLZ), Technische Universität München, 85748 Garching, Germany}

\author{Holger Kropf}
\affiliation{Helmholtz-Zentrum Berlin f\"ur Materialien und Energie, D-14109 Berlin, Germany}

\author{Moritz Winter}
\affiliation{Max Planck Institute for Chemical Physics of Solids, 01187 Dresden, Germany}
\affiliation{Dresden Center for Nanoanalysis, cfaed, Technical University Dresden, 01069 Dresden, Germany}

\author{Sebastian Schneider}
\author{Alexander Tahn}
\author{Bernd Rellinghaus}
\affiliation{Dresden Center for Nanoanalysis, cfaed, Technical University Dresden, 01069 Dresden, Germany}

\author{Tim A. Butcher}
\affiliation{Max Born Institute for Nonlinear Optics and Short Pulse Spectroscopy, 12489 Berlin, Germany}

\author{Simone Finizio}
\affiliation{Swiss Light Source, Paul Scherrer Institut, 5232 Villigen PSI, Switzerland}

\author{Sebastian Wintz}
\affiliation{Helmholtz-Zentrum Berlin f\"ur Materialien und Energie, D-12489 Berlin, Germany}

\author{Markus Weigand}
\affiliation{Helmholtz-Zentrum Berlin f\"ur Materialien und Energie, D-12489 Berlin, Germany}

\author{Max T. Birch}
\affiliation{RIKEN Center for Emergent Matter Science (CEMS), Wako 351-0198, Japan}

\author{Yoshinori Tokura}
\affiliation{RIKEN Center for Emergent Matter Science (CEMS), Wako 351-0198, Japan}
\affiliation{Department of Applied Physics, University of Tokyo, Tokyo 113-8656, Japan}
\affiliation{Tokyo College, University of Tokyo, Tokyo 113-8656, Japan}

\author{Yasujiro Taguchi}
\author{Kosuke Karube}
\affiliation{RIKEN Center for Emergent Matter Science (CEMS), Wako 351-0198, Japan}

\author{Florin Radu}
\affiliation{Helmholtz-Zentrum Berlin f\"ur Materialien und Energie, D-12489 Berlin, Germany}

\date{\today}

\begin{abstract}

(Fe$_{0.63}$Ni$_{0.3}$Pd$_{0.07}$)$_3$P is a room-temperature magnet with $S_4$ symmetry that hosts a rich variety of topological spin textures. Here, we report a combined resonant small-angle x-ray scattering and ptychography study of (Fe$_{0.63}$Ni$_{0.3}$Pd$_{0.07}$)$_3$P in vector magnetic fields over a broad temperature range. We demonstrate deterministic vector-field control of magnetic stripe domains, where in-plane fields continuously rotate their orientation via a transition from a chiral stripe to an achiral fan configuration. Furthermore, at 50\,K and below, the stripe orientation becomes metastably pinned and retains its field-trained direction. While the magnitude of the wavevector is nearly isotropic within the basal plane at room temperature, a pronounced temperature evolution of anisotropic interactions emerges upon cooling. In particular, non-trivial anisotropy axes develop at 20-50\,K reflecting the combined effects of magnetocrystalline anisotropy, anisotropic exchange, and Dzyaloshinskii-Moriya interaction (DMI), whose effective orientation is found to rotate with temperature. These results establish (Fe$_{0.63}$Ni$_{0.3}$Pd$_{0.07}$)$_3$P as a model system for vector-field control of chiral spin textures and reveal a previously unrecognized temperature-driven evolution of the effective DMI landscape in a noncentrosymmetric magnet.
\end{abstract}

\maketitle

\section{Introduction}

Noncentrosymmetric magnets with competing symmetric and antisymmetric exchange interactions provide fertile ground for the stabilization of complex spin textures such as helices, conical spirals, and skyrmion lattices \cite{bak1980theory,bogdanov1994thermodynamically}. In such systems, the absence of inversion symmetry allows for a Dzyaloshinskii–Moriya interaction (DMI) \cite{dzyaloshinsky1958thermodynamic,moriya1960anisotropic} that controls the rotational sense of the magnetization and competes with the exchange and anisotropy terms to produce long-periodic, modulated ground-state spin structures. The direction and symmetry of these modulations are essentially governed by the crystallographic class: for example, chiral cubic compounds host isotropic DMI leading to Bloch-type skyrmion lattices \cite{muhlbauer2009skyrmion,tokunaga2015new}, whereas polar, $D_{2d}$ and $S_4$ systems exhibit anisotropic Dzyaloshinskii–Moriya (DM) vectors that confine the propagation vectors to a high-symmetry crystallographic plane and stabilize N\'eel-type skyrmions \cite{kezsmarki2015neel,kurumaji2017neel} and antiskyrmions \cite{nayak2017magnetic,karube2021room}, respectively. However, the role of in-plane vector magnetic fields in controlling propagation vectors of magnetic textures in $S_4$-symmetric magnets remains poorly understood.

\begin{figure*}
\includegraphics[width=17cm]{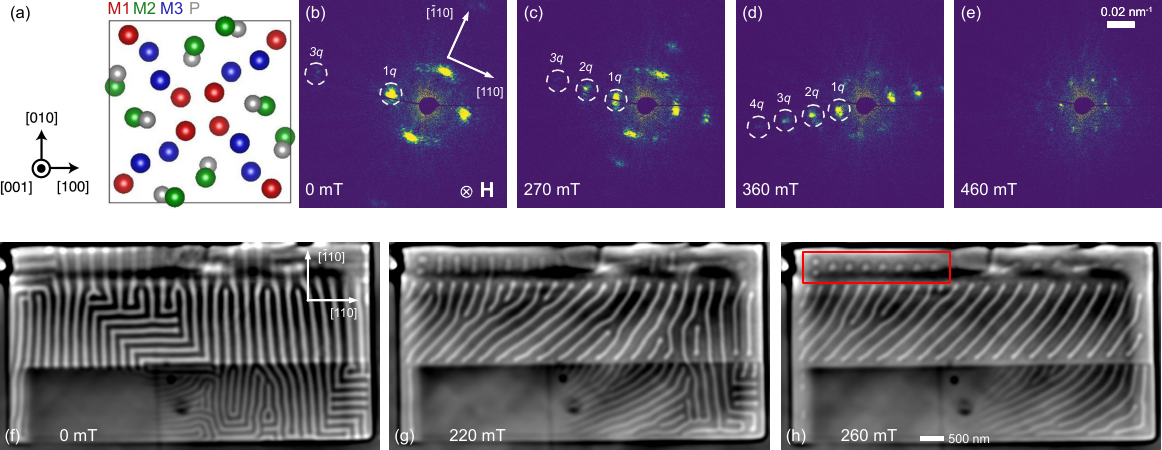}
\caption{(a) Schematic of noncentrosymmetric tetragonal crystal structure of (Fe$_{0.63}$Ni$_{0.3}$Pd$_{0.07}$)$_3$P viewed from [001] direction. Three inequivalent crystallographic sites, randomly occupied by metallic ions, are denoted M1, M2, and M3. (b-e) Resonant SAXS patterns measured in (b) zero field, (c) 270\,mT, (d) 360\,mT, and (e) 460\,mT applied along the $c$-axis. (f-h) Real-space ptychographic phase images of magnetic textures measured at the out-of-plane magnetic field $\mathbf{H}||[001]$ of (f) 0 mT, (g) 220 mT, (h) 260 mT in stripe, soliton, and co-existing soliton and skyrmion states. Skyrmions nucleated at 260 mT are highlighted by the red box. }
\label{fig1}
\end{figure*}

Recently, two families of tetragonal compounds have attracted attention for hosting room-temperature antiskyrmions. The first, which includes Heusler alloys Mn$_{1.4}$Pt$_{0.9}$Pd$_{0.1}$Sn and Mn$_{1.4}$PtSn (hereafter we only focus on the representative member Mn$_{1.4}$PtSn referred to as MPS), possesses $D_{2d}$ symmetry \cite{nayak2017magnetic}, while the second, Pd-doped schreibersites, represented by (Fe$_{0.63}$Ni$_{0.3}$Pd$_{0.07}$)$_3$P (hereafter FNPP), belongs to the $S_4$ class \cite{karube2021room,karube2022high}. FNPP crystallizes in a noncentrosymmetric structure with space group $I\bar{4}$, where Fe, Ni, and Pd atoms occupy three inequivalent Wyckoff positions, giving rise to anisotropic DMI confined within the basal plane. In contrast to $D_{2d}$, $S_{4}$ symmetry lacks mirror planes but retains a fourfold rotoinversion axis, resulting in a distinct set of DM vectors with mixed in-plane and out-of-plane components. In theory, this lower symmetry allows for more complex chiral modulations, including both Bloch-type helical and N\'eel-type cycloidal textures \cite{bogdanov1994thermodynamically,mukai2022skyrmion}. Additionally, in bulk MPS and FNPP, a strong uniaxial anisotropy and long-ranged magnetic dipole-dipole interaction \cite{ma2020tunable,karube2021room} dominate over DMI-induced topological spin textures and stabilize long-scale (tens of microns) fractal domain patterns in thick samples \cite{sukhanov2020anisotropic,karube2022unveiling}. We emphasize that due to the minor role of DMI in these two material systems, it is more appropriate to assign their ground state to a stripe domain phase \cite{hubert1998magnetic} rather than a helical spiral state typical to archetype chiral magnets such as MnSi \cite{bak1980theory}. In this context, the stripe-domain states realized in FNPP can be viewed as a limiting case of a $\pi$-rotation chiral soliton lattice ($\pi$-CSL) in the weak-DMI regime, providing a natural bridge between dipolar-driven magnetic domains and chiral soliton physics \cite{winter2025field}.

Unlike MPS, where the zero-field Bloch-type stripe domain texture propagates along the $\langle100\rangle$ axes, they orient preferably along the $\langle110\rangle$ diagonals in FNPP, due to the combined effects of easy-axis and basal anisotropies as well as the anisotropic DMI \cite{karube2022doping,hemmida2024role}.

Despite extensive investigations of the topology pertaining to the magnetic structures in these antiskyrmion hosts, key aspects related to detailed conditions for stabilization of diverse chiral magnetic structures remain unresolved. While the bulk magnetic properties of cubic chiral Bloch-type and polar N\'eel-type skyrmion hosts are well studied by small-angle neutron scattering (SANS) \cite{muhlbauer2019magnetic}, this approach is less effective for MPS and FNPP due to the large size of magnetic domains in bulk crystals \cite{sukhanov2020anisotropic,karube2022unveiling}. Consequently, most studies on these materials have relied on Lorentz transmission electron microscopy (LTEM) of thin lamellae. LTEM offers nanometric real-space imaging of magnetic textures, but it generally has limited options for changing the magnetic field direction, usually confined to tilt angles of 45 degrees or less \cite{peng2020controlled,jena2020elliptical,he2026manipulation}. Yet, the in-plane component of the external magnetic field, and its relative orientation to the in-plane crystal axes play an important role in stabilizing antiskyrmion spin textures in MPS and FNPP, by first nucleating so-called non-topological bubble state \cite{peng2020controlled,peng2022formation,tang2023sewing,muto2025switching,wu2025current}. 

\begin{figure*}
\includegraphics[width=17cm]{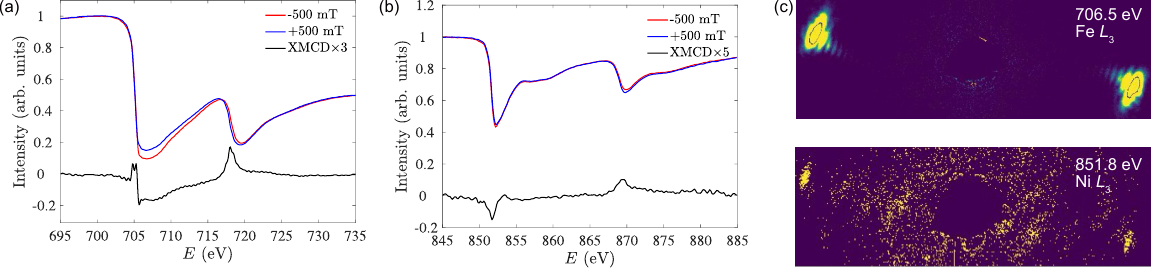}
\caption{(a,b) Soft x-ray transmission and XMCD spectra measured across Fe and Ni $L_{2,3}$ edges, respectively. (c) Rescaled element-selective SAXS patterns measured at zero field at Fe and Ni edges.}
\label{fig2}
\end{figure*}

Resonant small-angle X-ray scattering (SAXS) is a reciprocal-space technique that probes similar lamellae samples but offers a more flexible magnetic field environment \cite{baral2023direct}. Modern synchrotron-based setups provide full three-dimensional vector-field control, enabling studies in geometries inaccessible to LTEM. Previous works have largely examined FNPP lamellae aligned along the $c$ axis or in perpendicular and tilted fields, leaving the response to purely in-plane fields unexplored. Moreover, the interplay between basal magnetocrystalline anisotropy, anisotropic exchange interaction (AEI) \cite{ukleev2021signature}, and the anisotropic DMI inherent to the $S_4$ symmetry has not yet been addressed.

Here, we employ element-selective resonant SAXS under vector magnetic fields and x-ray ptychography to address these open questions. By applying in-plane vector fields of controlled magnitude and direction within the (001) plane, we map the orientation and magnitude of the magnetic modulation vector $\mathbf{q}$ in FNPP over a broad temperature range (20-300\,K) using SAXS. We find that at room temperature, the propagation vector of magnetic stripe domains can be continuously rotated by weak in-plane fields of about 10\,mT, demonstrating full vectorial control of stripe orientation within the basal plane. In this regime, the Zeeman energy competes with and eventually overcomes the Dzyaloshinskii–Moriya interaction that stabilizes the chiral modulation, driving a transition from stripe domains to an achiral fan state. Upon cooling to 50\,K and below, a weak fourfold anisotropy emerges, leading to spontaneous locking of the modulation direction to the field-training axis even after field removal. The temperature evolution is quantitatively reproduced by a theoretical model incorporating DMI, magnetocrystalline square anisotropy, and anisotropic exchange, which together establish an effective temperature-dependent basal-plane anisotropy governing the stripe-domain orientation.

\section{Results and Discussion}

\subsection{Magnetic textures in zero and out-of-plane fields}

FNPP crystallizes in tetragonal structure with the unit cell ($a=9.1201(6)$\,\AA~and $c=4.4907(3)$\,\AA) as schematically shown in Fig.~\ref{fig1}a with the ground state magnetic modulation confined to the $\langle110\rangle$ crystal axes \cite{karube2021room}. SAXS patterns measured at zero field show that the stripe domain state manifests as four magnetic Bragg peaks corresponding to two symmetry-equivalent directions (Fig.~\ref{fig1}b). The real-space period is $\lambda_0 = 234 \pm 3$~nm.  Notably, even at zero field, SAXS reveals not only the first-order harmonics ($1q$) but also third-order ones ($3q$), while the second order is suppressed. Similar zero-field x-ray scattering patterns featuring third-order harmonics were also observed in MPS \cite{winter2025field}, Co/Pt multilayers \cite{hellwig2003x}, and from surface of Cu$_2$OSeO$_3$ \cite{baral2026surface}. The observation of pronounced third-order scattering indicates a strongly non-sinusoidal stripe (or $\pi$-CSL) modulation, consistent with sharp domain walls observed in the real-space ptychographic images (Fig. \ref{fig1}f), which can be viewed as a distorted square-wave-like magnetic profile.

The slight departure of one of the zero-field stripe domains’ orientation away from the crystal [$\bar{1}$10] axis is, most likely, influenced by the edge of the lamella (see Fig. S1 of the Supplementary Note I). Upon increasing the magnetic field parallel to the $c$ axis, SAXS reveals transitions from a multidomain chiral stripe-domain state (Fig.~\ref{fig1}b) into a soliton-lattice state with multiple higher harmonics up to 4$^{\mathrm{th}}$ order (Fig.~\ref{fig1}c,d). Finally, a single-domain hexagonal skyrmion lattice (Fig.~\ref{fig1}e) is formed which is evidenced by a characteristic sixfold scattering pattern with weaker higher harmonics \cite{adams2011long,moskvin2013complex,ukleev2020metastable,reim2022higher}. Above $\sim$480~mT, the twisted texture unwinds, and the SAXS signal vanishes. Just before saturation, the modulation period increases by $\sim$30\% with respect to zero field ($\lambda = 340 \pm 3$~nm). Ptychographic phase images reconstructed from the SAXS signal of another FNPP lamella with distinct thickness steps from $\sim 50$\,nm to $\sim 200$\,nm are shown in Figs.~\ref{fig1}f-h. Stripes, solitons, and first skyrmion nucleation in the thicker part of the lamella are clearly reproduced. The increase of the period of magnetic textures with thickness and the sequence of transitions, correspond well with the magnetic phase diagram obtained previously by LTEM \cite{karube2021room}.

\begin{figure*}
\includegraphics[width=17cm]{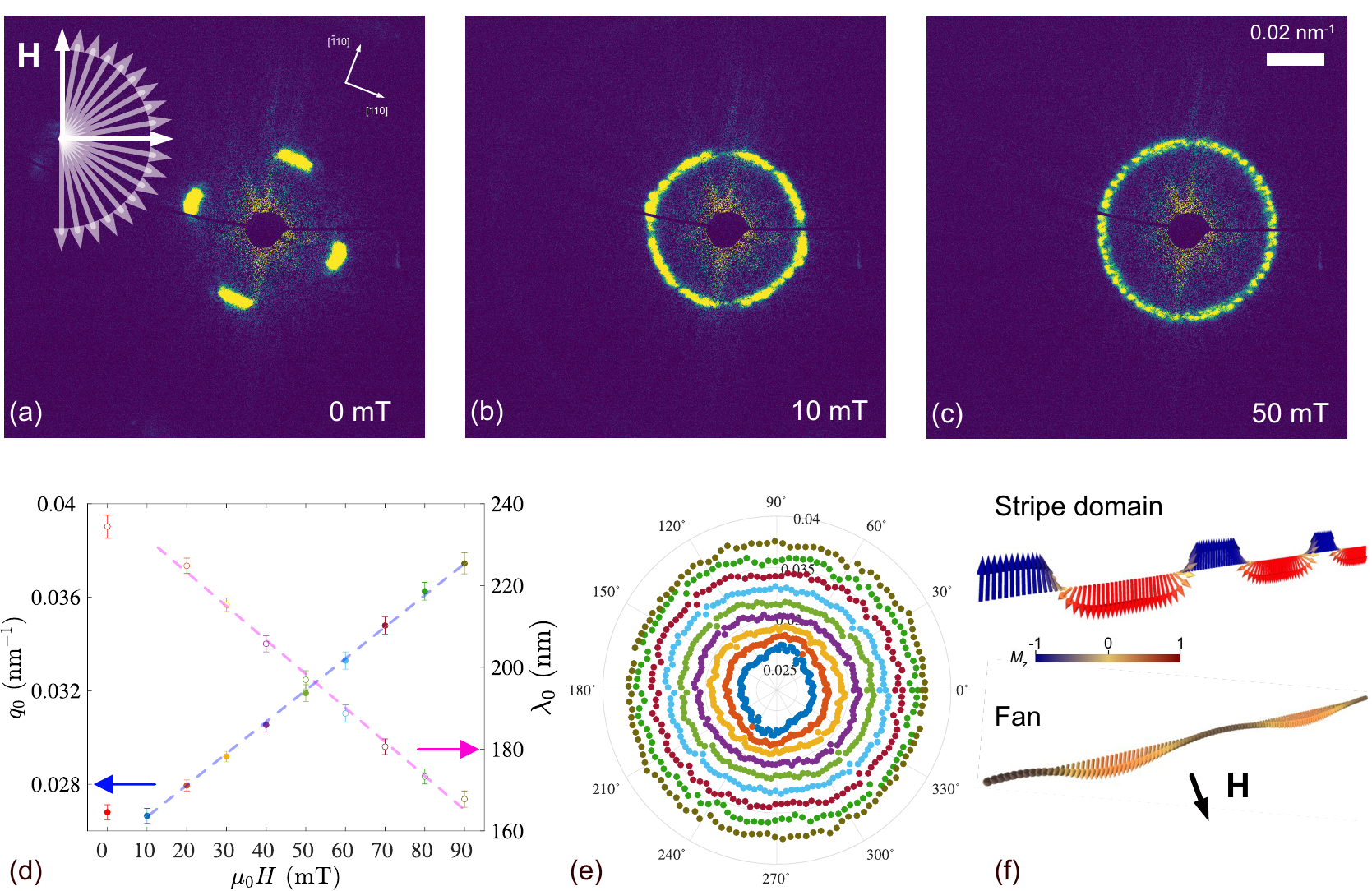}
\caption{Summed vector-field SAXS patterns measured at 300\,K and at the magnitude of the azimuthal magnetic field of (a) 0\,mT, (b) 10\,mT, and (c) 50\,mT. (d) In-plane field dependence of the magnetic wavevector and the corresponding real-space stripe periodicity averaged over all azimuthal field angles. (e) Polar plot of the wavevector magnitude as a function of the azimuthal field angle and the magnitude of the field. Symbol colors correspond to those indicating the field magnitudes in panel (d). (f) Schematics of zero-field stripe domain and in-plane field-induced fan structures.}
\label{fig3}
\end{figure*}

\subsection{Element-selective spectroscopy}

By tuning the photon energy to the $L_{2,3}$ absorption edges of different transition metals, one can obtain element-selective information on the magnetization of each sublattice. Here, X-ray magnetic circular dichroism (XMCD) and SAXS were used to probe the magnetic ordering of Fe and Ni in FNPP, independently.
 
XMCD spectra (Figs.~\ref{fig2}a,b) measured in transmission geometry in the field-polarized state at the field $\mu_0 H = 500$\,mT applied along the $c$-axis show strong magnetic contrast at both Fe and Ni $L_{2,3}$ edges. The same signs of the XMCD signal for both elements indicate parallel alignment of their moments. The presence of finite XMCD at the Ni edge is remarkable, as Ni$_3$P compound is paramagnetic \cite{gambino1967magnetic}, i.e., Ni carries only field-induced, spatially uncorrelated magnetic moments that do not participate in the long-range modulated magnetic structure. However, element-resolved SAXS maps (Fig.~\ref{fig2}c), measured at zero field, display coincident magnetic Bragg peaks at the same finite wavevector $q$ at both the Fe (706.5\,eV) and Ni (851.3\,eV) resonances. A purely paramagnetic contribution would result only in a uniform ($q=0$) magnetization and cannot produce scattering at finite $q$. The observation of identical scattering patterns therefore demonstrates that the Ni moments are spatially modulated and phase-locked to the Fe magnetic structure. This rules out a purely paramagnetic origin of the XMCD signal at the Ni edge and instead indicates that Ni carries an induced magnetic moment via proximity to Fe within the magnetically ordered state.

All subsequent SAXS measurements were performed at the Fe $L_3$ edge (706.5\,eV), where the magnetic satellite intensity is maximized.

\subsection{Vector-field control of stripe domain orientation}

A key advantage of SAXS is the availability of a vector magnetic field at VEKMAG. In previous studies, the magnetic wavevector control in MPS by an $ex-situ$ in-plane field training \cite{sukhanov2022hybrid}, and in FNPP by a mechanical strain \cite{mori2025direct} has been demonstrated. Vector-field SAXS provides direct insights into the magnetic domain anisotropy. The response of the stripes to in-plane magnetic fields applied at an in-plane angle $\theta$ is shown in Fig.~\ref{fig3}. SAXS patterns were recorded while rotating the in-plane magnetic field azimuthally in $3^\circ$ steps; $\theta=0$ corresponds to the vertical direction in the laboratory reference system.  

\begin{figure*}
\includegraphics[width=17cm]{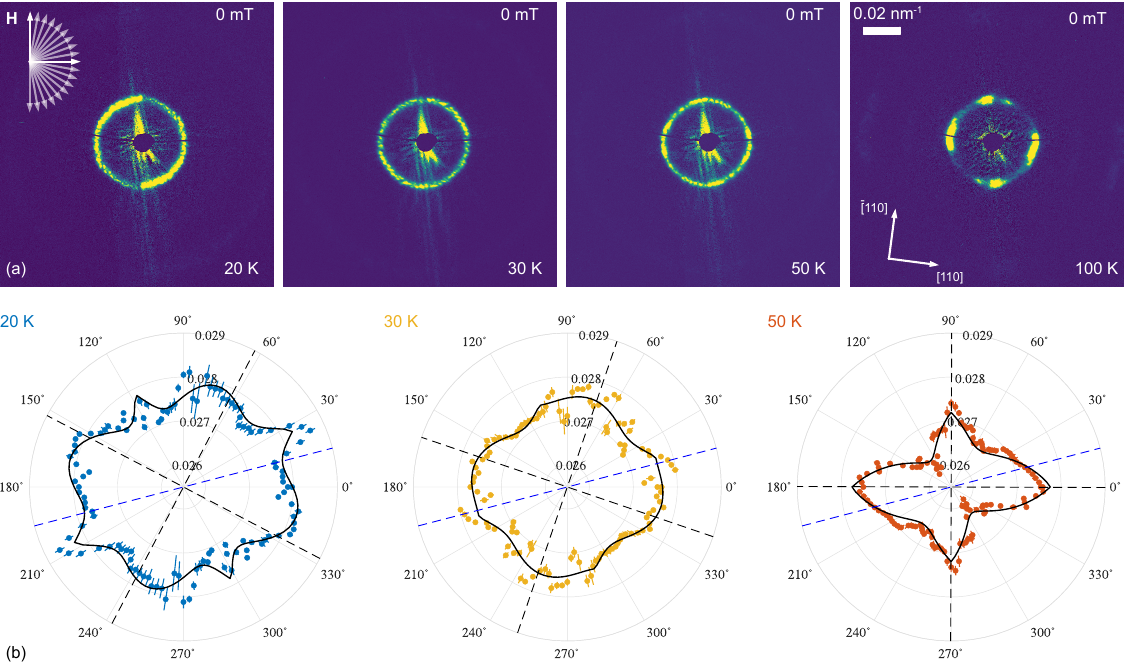}
\caption{(a) Summed zero-field SAXS patterns measured at 20\,K, 30\,K, 50\,K, 100\,K after applying a saturating field of 200\,mT in every azimuthal direction with a 3$^\circ$ step. At 100\,K and above, the domain propagation direction always returns to its original axis. (b) Polar plot of the magnetic wavevector magnitude as a function of the azimuthal field angle at 20\,K, 30\,K, and 50\,K at zero field. Solid curves represent the fit according to the theoretical model discussed in the main text. Dashed black lines indicate the fitted orientations of the effective DMI axes, while the dashed blue line marks the direction of the in-plane uniaxial strain.}
\label{fig4}
\end{figure*}

Individual SAXS patterns for the in-plane field $\mathbf{H}\parallel[100]$ crystal axis are shown in the Supplementary Note II. Summing the individual azimuthal SAXS maps measured at 300\,K for field amplitudes of 0, 10, and 50\,mT results in the patterns shown in Figs.~\ref{fig3}a–c. The averaged SAXS pattern at zero magnetic field reveals that the magnetic propagation vectors are pinned along [110] and [$\bar{1}$10] due to anisotropic DMI favoring the Bloch domain walls in these directions. N\'eel structures, which can in principle emerge in [100] and [010] directions, are strongly suppressed by the magnetostatic energy (see Supplementary Note III for details). A weak in-plane field of 10~mT collapses the intensity into a pair of spots perpendicular to the field, indicating that $\mathbf{q}$ aligns normal to $\mathbf{H}$. Thus, even a weak external magnetic field of 10~mT is sufficient to align the domains in any in-plane direction, overcoming the anisotropic effect of DMI. This also indicates weak basal-plane square anisotropy and anisotropic exchange.

Upon further increasing the in-plane field, the system undergoes a transition from the chiral stripe state to an achiral fan-like configuration, as illustrated in Fig.~\ref{fig3}f. This transition occurs when the Zeeman energy gain of the fan state overcomes the DMI-induced chiral twisting (see Supplementary Note III for details). The corresponding critical field is given by
\begin{equation}
  H_\textrm{C} =  \frac{M_\textrm{s}\gamma}{\sqrt{\alpha/\beta}},
\end{equation}
where $M_\textrm{s}$ is the magnetization modulus at a given temperature, $\alpha$ is the exchange stiffness, $\beta$ -- perpendicular  anisotropy, and $\gamma$ is the effective DMI strength. Evidently, $\sqrt{\alpha/\beta}$ corresponds to the well-known domain wall width. Using parameter values from Ref.~\cite{karube2021room}, we estimate a transition field of approximately $30$\,mT (see Supplementary Note III). Above this field, and in the absence of significant basal-plane anisotropy, the modulation wavevector becomes isotropic, as the influence of DMI on its orientation is effectively suppressed. This behavior is consistent with our experimental observations shown in Fig.~\ref{fig3}e.

Increasing the magnetic field amplitude to 90\,mT reduces the stripe period to $\lambda_0 = 167 \pm 3$~nm (see Fig.~\ref{fig3}d), in contrast to the $\mathbf{H}\parallel[001]$ geometry, and consistent with systems combining DMI and easy-axis anisotropy \cite{yang2021universality}. After removing the field, the original domain orientations are restored, with $\mathbf{q}$ vectors pinned along the $\langle110\rangle$ directions, analogous to the behavior in Cu$_2$OSeO$_3$ under vectorial fields \cite{moody2021experimental,baral2023direct}.

Polar plots of $\mathbf{q}$ (Fig. \ref{fig3}e) show a continuous rotation of the stripe orientation with the azimuthal field direction. At room temperature, the $\mathbf{q}$ distribution is isotropic in the (001) plane and reflects neither the crystal symmetry nor the sample shape. In contrast to cubic chiral magnets, where the symmetry-allowed conical state constrains the field response, in FNPP the absence of such a conical phase permits an arbitrary reorientation of the in-plane modulation vector via a fan state. 

%A weak in-plane magnetic field drives the system from the chiral stripe domain state into an achiral fan texture with $\mathbf{q}$ aligned perpendicular to the field (Fig. \ref{fig3}f). The fact that a weak field of 10\,mT is sufficient to destabilize the chiral stripe state and generate the fan texture further highlights the intrinsically weak DMI energy scale in FNPP. Corresponding micromagnetic simulations for the field-aligned stripe and fan states are shown in the Supplementary Information (Fig. S. 1).

To summarize, at room temperature, we observe the following scenario for the in-plane field-induced transitions: in the range of magnetic fields $<10$~mT, related to the scale of the anisotropic DMI, chiral Bloch stripe structures with pinned directions are stabilized. Weak magnetic fields $\gtrsim 10$~mT lead to orientation of the stripe $\mathbf{q}$-vectors perpendicular to the field direction and transform them into nonchiral fan structures, see Fig.~\ref{fig3}f. Notably, in this regime, the effective role of anisotropic DMI in forming the chiral domain texture is suppressed, and basal-plane anisotropy is negligible; hence, the stripe domain structure period is nearly isotropic. Upon further in-plane field increase, the domain wall energy drops down~\cite{yang2021universality} and the period decreases (see also Supplementary Note III). The stripe domains progressively transition into a fan-like structure. This transformation scenario is also supported by micromagnetic simulations presented in the Supplementary Information (see Fig.~S.~4). A comprehensive micromagnetic study into the mechanism underlying the stripe-to-fan transformation is presented in a separate study on MPS
 \cite{winter2026quasi1dspintextureschiral}.
 
\subsection{Temperature evolution and basal-plane anisotropy}

Upon cooling, the vectorial magnetic field control of $\mathbf{q}$ changes drastically. At 50\,K, zero-field alignment of domains by field training becomes possible. Figure~\ref{fig4}(a) presents SAXS intensity distributions at zero field after applying and removing an in-plane field at various temperatures. At 100\,K and above, $\mathbf{q}$ reverts to $\langle110\rangle$ after field removal, whereas at 50\,K and below the stripes retain the trained orientation, revealing metastable directional memory. This behavior resembles directional pinning of helices observed in chiral FeGe \cite{ukleev2021signature} and Co$_8$Zn$_8$Mn$_4$ \cite{ukleev2024competing}, and is similar to the low-temperature state in MPS below the spin-reorientation transition \cite{sukhanov2022hybrid}.

The energy of Bloch-type stripe domains depends on their orientation through anisotropic DMI and basal-plane anisotropy, leading to an angular variation of the stripe wavevector $q(\theta)$. Polar plots of $q(\theta)$ (Fig.~\ref{fig4}b) reveal a clear fourfold anisotropy at 50\,K, while more complex shapes are observed at lower temperatures.

To quantify this behavior, we model $q(\theta)$ using a phenomenological approach that perturbatively incorporates the DMI, an effective basal-plane anisotropy, and a strain-induced in-plane uniaxial anisotropy. The effective basal anisotropy combines magnetocrystalline square anisotropy and anisotropic exchange interaction (AEI), which cannot be disentangled within the present experimental sensitivity. The uniaxial term arises naturally from the lamella preparation by Focused Ion Beam (FIB), where the sample is mounted onto a silicon nitride membrane using Pt contacts (see Fig.~S.1c). This introduces residual strain, which is further enhanced upon cooling due to thermal expansion mismatch and is known to influence magnetic textures. The role of strain in chiral magnets has been widely demonstrated, including x-ray studies on Cu$_2$OSeO$_3$ \cite{okamura2017emergence} and FeGe \cite{ukleev2020metastable}, as well as LTEM investigations of Co$_8$Zn$_{8.5}$Mn$_{3.5}$ \cite{zhao2025controlling} and FNPP \cite{mori2025direct}. In FNPP specifically, strain has been shown to modify stripe-domain orientation even at zero field \cite{mori2025direct}.

Within this framework, the characteristic wavevector can be expressed as

\be \label{eq1}
\begin{aligned}
 q(\theta) = q_0 \Big[ 1 + b \sin^2{(\theta-\theta_\textrm{S})}  
 + d |\cos{ 2 (\theta-\theta_\textrm{D})}| \\
   + f \sin^2{2\theta} \Big]
\end{aligned}
\ee
where $b \sim \beta_\textrm{S}/ \beta$ accounts for the strain-induced anisotropy, $d \sim \gamma/\sqrt{\alpha \beta}$ quantifies the DMI contribution, $f$ represents the effective basal anisotropy (see Supplementary Note III). Here, $\gamma$ is the DMI constant, $\alpha$ is the exchange stiffness, $\beta$ and $\beta_\textrm{S}$ are for perpendicular and in-plane strain-induced axial anisotropies. Angles $\theta_\textrm{D}$ and $\theta_\textrm{S}$ indicate the orientations of the DMI and strain axes, respectively.

For the 20\,K data, the fitted model yields $d \approx 0.073 \pm 0.019$ and $b \approx -0.023 \pm 0.015$. These parameters are nearly temperature-independent and remain within the experimental uncertainty. In contrast, the effective anisotropy parameter $f$ increases by an order of magnitude upon cooling from 50\,K to 20\,K, changing from negligible $f = 0.001 \pm 0.006$ to significant $f = -0.070 \pm 0.015$. This strong enhancement is consistent with previous ferromagnetic resonance measurements on bulk FNPP \cite{hemmida2024role} and general expectations, since higher-order magnetocrystalline anisotropy scales with higher powers of magnetization and anisotropic exchange becomes more prominent as thermal fluctuations are suppressed \cite{baral2023direct,ukleev2024competing}. In addition, the random occupation of Fe, Ni, and Pd sites, combined with the strong spin–orbit coupling of Pd, may give rise to locally varying anisotropies that become increasingly relevant at low temperatures. Moreover, the contribution of the square magnetocrystalline anisotropy scales with the square of the magnetization, leading to a temperature-dependent correction of order $\sim (T/T_\textrm{C})^{3/2}$ for $T \ll T_\textrm{C}$. This implies that the conventional enhancement of the basal-plane anisotropy due to the increase of the saturated magnetization upon cooling from $50$~K to $20$~K is limited to approximately $10\%$, and therefore cannot account for the pronounced experimentally observed growth. In contrast, a substantial increase of the AEI at temperatures well below the ordering temperature has been reported for cubic helimagnets \cite{baral2023direct,ukleev2021signature}, suggesting that AEI provides a more plausible microscopic origin for the strong low-temperature enhancement of the basal anisotropy.

At 50\,K, where the basal anisotropy remains relatively weak, the DMI term produces a cusp-like angular dependence of $q(\theta)$ near the $\langle110\rangle$ directions, while the strain-induced contribution introduces a small asymmetry between otherwise equivalent axes (the strain direction is indicated in Fig.~\ref{fig4}b). Upon further cooling, we observe a systematic rotation of the effective DMI axis $\theta_\textrm{D}$ by approximately $20^\circ$ at 30\,K and $30^\circ$ at 20\,K. Such a rotation is not symmetry-forbidden for the $S_4$ group and points to a temperature-dependent renormalization of the DMI tensor. Microscopically, this may originate from a weak symmetry lowering within the basal plane-e.g., toward an orthorhombic distortion—driven by magnetoelastic coupling or differential thermal contraction, which modifies the underlying spin–orbit–coupled exchange paths. In addition, the progressive freezing of spin and phonon fluctuations can further renormalize the balance of competing anisotropic interactions. These results provide experimental evidence that in complex tetragonal magnets the effective DMI orientation need not be rigidly symmetry-locked, but can evolve with temperature. A quantitative understanding of this effect will require microscopic modelling, for instance based on temperature-dependent \textit{ab initio} calculations of the anisotropic interactions.

Overall, the evolution from nearly isotropic behavior at high temperatures to pronounced fourfold (50-30\,K) and eventually more complex anisotropy at 20\,K reflects a temperature-dependent competition between DMI, strain, and basal-plane anisotropies. While DMI governs the preferred modulation directions at elevated temperatures, the increasing strength of magnetocrystalline and exchange anisotropies at low temperatures distorts the angular dependence and stabilizes specific domain orientations.

\section{Conclusion}

We have investigated magnetic stripe domains in noncentrosymmetric (Fe$_{0.63}$Ni$_{0.3}$Pd$_{0.07}$)$_3$P using element-selective resonant soft x-ray scattering and spectroscopy under vector magnetic fields, in addition to imaging topological magnetic textures in real space by soft x-ray ptychography. The element selectivity of soft x-rays reveals an induced magnetic moment in Ni, which is absent in the parent compound Ni$_3$P, confirming a strong exchange coupling between the Fe and Ni sublattices. At room temperature, the magnetic modulation vector can be continuously rotated by weak in-plane fields of only a few millitesla, demonstrating nearly isotropic modulation within the basal plane. Upon cooling below 50\,K, a fourfold basal-plane anisotropy emerges, and the stripe-domain orientation becomes metastably pinned along the direction of the previously applied field. Notably, we observed and quantified the temperature evolution of effective DMI orientation caused by a competition of two symmetry-allowed contributions of different origin.

Furthermore, we develop a closed analytical description of the stripe-to-fan transition under in-plane fields that incorporates anisotropic DMI together with basal anisotropy, yielding a quantitative expression for the critical field that governs the stability and reorientation of chiral stripe domains. This provides the first explicit analytical framework linking DMI, anisotropic exchange, and vector-field-driven domain reconfiguration in tetragonal noncentrosymmetric magnets.

Our results show that FNPP hosts a competition between nearly isotropic and anisotropic magnetic interactions, enabling vector-field control of magnetic domain orientation and directional memory effects at cryogenic temperatures. Rather than establishing FNPP as a generic model system, our findings identify a concrete mechanism by which weak in-plane fields reorganize chiral stripe states through a DMI-anisotropy-Zeeman balance. More broadly, these results demonstrate how vector magnetic fields can be used to manipulate mesoscale magnetic textures in noncentrosymmetric magnets, providing a route toward controllable topological spin structures in functional magnetic materials.

\section{Experimental section}

\subsection{Samples}
Single crystals of (Fe$_{0.63}$Ni$_{0.3}$Pd$_{0.07}$)$_3$P were grown by a self-flux method and characterized as described previously \cite{karube2021room}. The crystals were oriented by Laue diffraction, and a thin ($\sim$120~nm) (001)-oriented lamella for transmission x-ray measurements was prepared by focused ion beam (FIB) milling. The lamella was attached to a masked Si$_3$N$_4$ membrane (Silson Ltd., UK) with a 5~$\mu$m circular pinhole using two Pt contacts. Details of similar sample preparation routines and associated challenges are discussed in Refs.~\cite{ukleev2019element,ukleev2020metastable}. LTEM and scanning electron microscopy (SEM) images of the lamella before and after its deposition on a membrane are shown in the Supplementary Note I.

\subsection{Resonant small-angle x-ray scattering and spectroscopy}
Resonant SAXS measurements in vectorial magnetic fields were performed at the PM-2 VEKMAG \cite{noll2016mechanics}, and preliminary SAXS data was acquired at PM-3 ALICE II endstation of the BESSY~II synchrotron (Berlin, Germany). At both instruments circularly polarized soft x-ray beam with the size of 100 $\mu$ were tuned to the Fe and Ni $L_{3}$ edges. The FNPP lamella was oriented perpendicular to the x-ray beam. Details of the VEKMAG setup for SAXS measurements are provided in Refs.~\cite{baral2023direct,ukleev2021signature,ukleev2024competing}. In particular, VEKMAG allows for soft x-ray spectroscopy and SAXS measurements in vectorial magnetic field: a 9/2/1~T cryomagnet equipped with three superconducting coils provides independent control of the magnetic field components, resulting in up to 1~T in any spatial direction. The acquisition time per SAXS pattern was 20 seconds. An image containing only the charge scattering was taken at the saturating field of 0.5\,T, and subtracted as background.

Sample alignment and energy-dependent x-ray spectra were recorded using a transmission photodiode. X-ray magnetic circular dichroism (XMCD) data were collected in oppositely saturated magnetic states at $\mu_0 H = \pm0.5$~T in normal geometry.

\subsection{Soft x-ray ptychography}
Soft x-ray ptychography measurements \cite{butcher2025soft}, which allow nanoscale imaging of noncollinear magnetic order by reconstruction of resonant x-ray scattering \cite{butcher2024ptychographic}, were carried out at the MAXYMUS microscope at BESSY II. The experiment was performed at room temperature in an out-of-plane magnetic field up to 260\,mT provided by a system based on permanent magnets \cite{nolle2012note}. Ptychographic images were obtained at the Fe $L_3$ pre-edge (1\,eV below the absorption maximum to enhance the phase contrast) with circularly polarized x-rays, using an inverse low-gain avalanche diode (iLGAD) Eiger soft-x-ray detector with $512\times512$ pixels (75\,$\mu$m pixel size) \cite{baruffaldi2025single}. The distance between the sample and the detector was 76\,mm. An exposure time of 200\,ms was used to record diffraction patterns from 100\,nm spaced areas illuminated by a 900\,nm full width at half maximum (FWHM) x-ray spot, which was shaped by a 320\,$\mu$m diameter Fresnel zone plate with 18\,nm outer zone width. Ptychographic reconstructions were carried out with the difference-map \cite{thibault2008high} and maximum-likelihood methods \cite{odstrvcil2018iterative}
as implemented in the PtychoShelves software \cite{wakonig2020ptychoshelves}.

\section*{Acknowledgments}

REXS experiments were carried out the BESSY II synchrotron as a part of proposals Photons-232-12305-ST, 232-12260-ST and 242-12606-EF. We thank the Helmholtz-Zentrum Berlin f\"ur Materialien und Energie for the allocation of beamtime. We thank R. Zehmke and T. Kachel for technical support. We thank M. Rahn, A. Sukhanov and J. Masell for fruitful discussions.

We thank the CoreLab Correlative Microscopy and Spectroscopy at Helmholtz-Zentrum Berlin, and Dresden Center for Nanoanalysis for the sample preparation using FIB.

We thank E. Fröjdh, F. Baruffaldi, M. Carulla, J. Zhang, and A. Bergamaschi of the Detector Group of the Paul Scherrer Institute (Villigen, Switzerland) Center for Photon Science for the development and provision of the iLGAD EIGER detector. The iLGAD sensors were fabricated at Fondazione Bruno Kessler (Trento, Italy).

The authors acknowledge financial support for the VEKMAG project and for the PM2-VEKMAG beamline by the German Federal Ministry for Education and Research (BMBF 05K2010, 05K2013, 05K2016, 05K2019) and by HZB. The study was also supported by JST CREST (Grant No. JPMJCR20T1), and the RIKEN TRIP initiative (Many-body Electron Systems and Advanced General Intelligence for Science Program). F.R. acknowledges funding by the German Research Foundation via Project No. SPP2137/RA 3570. O.I.U. acknowledges financial support from the Brain Pool Plus Program through the National Research Foundation of Korea funded by the Ministry of Science and ICT (2020H1D3A2A03099291). T.A.B. acknowledges funding from the European Regional Development Fund (ERDF). K.K. acknowledges financial support from the Japan Society for the Promotion of Science (KAKENHI, Grant No. 23K26534) and from the Japan Science and Technology Agency through the FOREST Program (Grant No. JPMJFR235R).

\section*{Author contributions}

K.K., Y. To., Y. Ta. prepared and characterized single crystals; V.U., L.U., C.L., R.A., P.W., F.R. performed x-ray scattering measurements; V.U., C.L., T.A.B., S.F., S.W., M.We. carried out x-ray ptychography measurements, V.U., H.K., M.Wi., S.S., A.T., B.R., M.B. prepared lamellae samples; O.I.U. and V.U. developed the theory and performed simulations; V.U., O.I.U. wrote the manuscript; V.U., K.K. and F.R. jointly conceived the project.

\section*{Conflict of interest}
The authors declare no conflict of interest.

\section*{Data availability statement}
The data that support the findings of this study are available from the corresponding author upon reasonable request.

\bibliography{biblio}
\end{document}